\documentclass[12pt]{iopart}

\usepackage{graphicx}
\usepackage{epsf}
\usepackage{epsfig}
\usepackage[figuresright]{rotating}

\newcommand{\be}{\begin{eqnarray}}

\newcommand{\ee}{\end{eqnarray}}

\begin{document}

\title{Theory Summary: Quark Matter 2006}
\author{Larry McLerran}
\address{Physics Department and Riken Brookhaven Center,
PO Box 5000, Brookhaven National Laboratory,  
 Upton, NY 11973 USA}

\begin{abstract}
I report on the theoretical developments discussed in Quark Matter 2006. 
\end{abstract}

\section{Introduction}

In Fig. 1, I show Hatsuda's artistic rendering of the relationship between the big bang and
ultra-relativistic heavy ion collisions.  The Color Glass Condensate
describes the very early stages of the collisions.  This is similar to the initial singularity problem
in cosmology. Fluctuations generated at early times eventually appear in the black body radiation
spectra and ultimately in the distribution of galaxies and clusters of galaxies In cosmology, the matter rapidly equilibrates, analogous in heavy ion collisions
to the formation and equilibration of a Quark Gluon Plasma.  In cosmology there are a variety of phase transitions which occur at later times, corresponding to the confinement-deconfinement transition
of QCD.    This analogy between ultra-relativistic heavy ion collisions, the Little Bang, and the Big Bang is deep, and I will expand on it later.  The main reason I show the figure is simply because it is art,
it is beautiful, and mixes together concrete realizations of science with those of the subjective imagination.

 In Fig. 2, I show Bass's famous artistic rendering of heavy ion collisions.  The careful reader
\begin{figure}[ht]
    \begin{center}
        \includegraphics[width=0.60\textwidth]{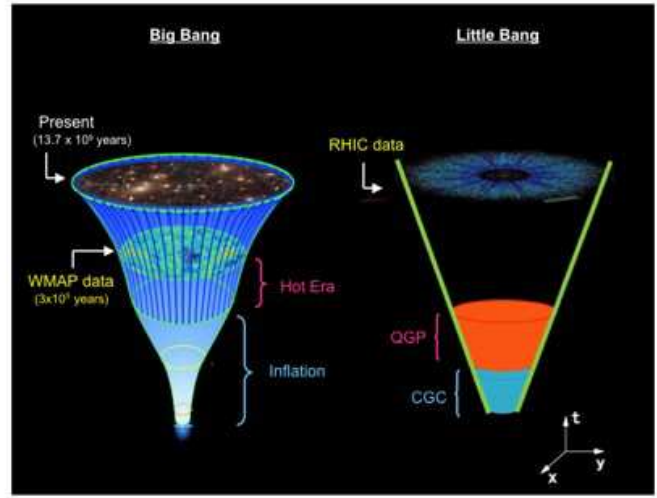}
        \caption{Hatsuda's picture of the relationship between heavy
        ion collisions and cosmology. }
\label{hatsuda}
    \end{center}
\end{figure}
\begin{figure}[ht]
    \begin{center}
        \includegraphics[width=0.80\textwidth]{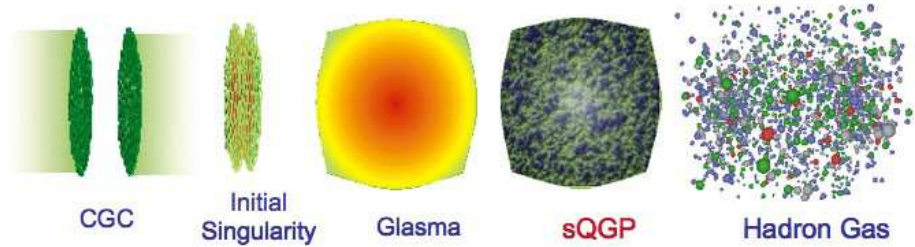}
        \caption{Stephen Bass's artistic rendering of the space-time evolution of heavy ion collisions. }
\label{bass}
    \end{center}
\end{figure}
will see that the labels have been changed from those used by Bass.  The labels still are appropriate,
and relate directly to the figure of Hatsuda, and what I shall discuss later.  Like good art,
it adapts to different trends, and changing ways of conceptualizing our knowledge.  
We need more art in physics.  It is an effective way to communicate the ideas which drive us, symbolizing our goals, accomplishments and aspirations. 

Much of the theoretical discussion at this meeting centered around the development of 
a  complete description of the interactions of hadrons at high energy, and their subsequent evolution
into a thermalized Quark Gluon Plasma.  The issues are important for Quark Gluon Plasma physics,
and more generally for the our understanding of the high energy limit of strong interactions.
We have heard about the Color Glass Condensate which is
the high density gluonic matter which describes the wee parton part of the wavefuction of a very
high energy hadron.  We have seen how the Glasma is produced almost infinitesimally after
the collisions of two high energy hadrons, and how the high energy density fields
of the Glasma evolve, and perhaps thermalize or perhaps make a turbulent fluid.  We have seen theoretical descriptions of the thermalized Quark Gluon Plasma, and its remarkable properties
of low viscosity and high opacity.  

There is a new paradigm:  Various forms of matter control
the high energy limit of QCD.  These forms of matter have intrinsically interesting properties.
This is the traditional  domain of nuclear physics and is at the basis of our methodology for
understanding nature.  Such a methodology is phenomenally successful and we should be proud of its accomplishments.

\begin{figure}[ht]
    \begin{center}
        \includegraphics[width=0.80\textwidth]{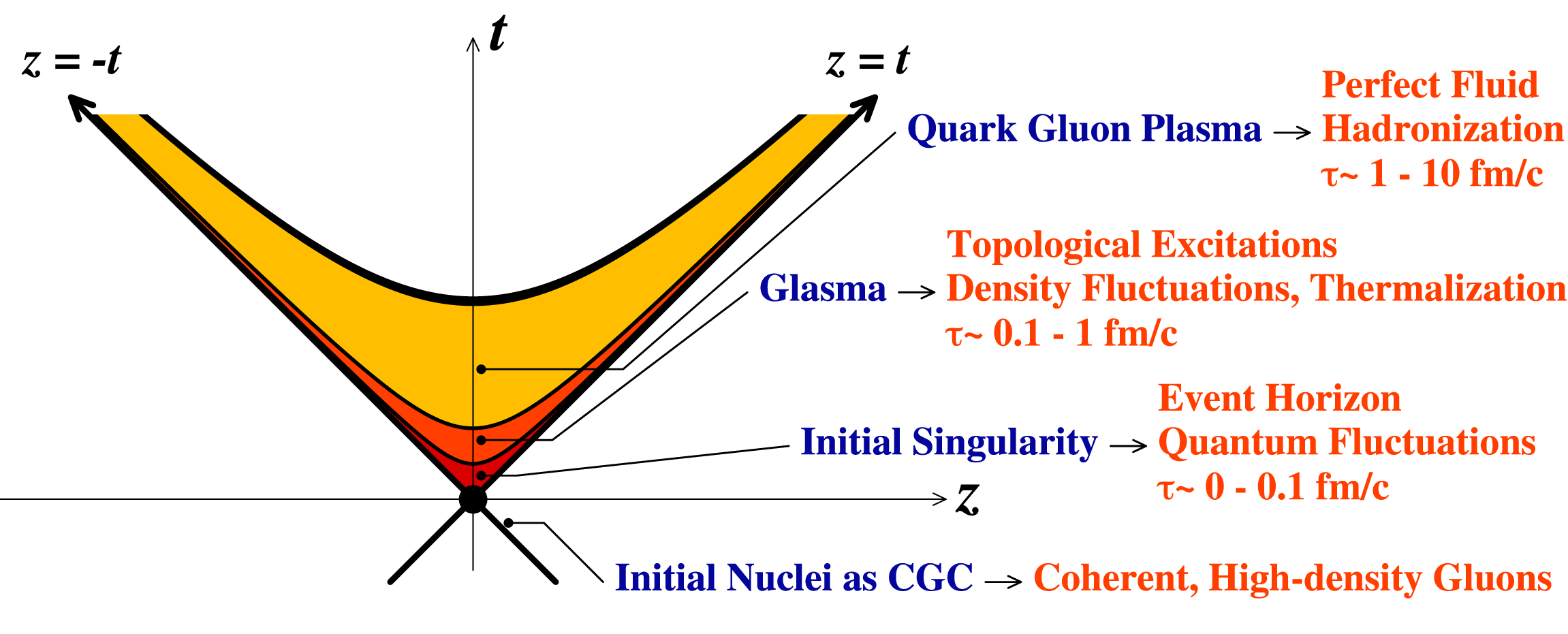}
        \caption{A schematic picture of the evolution of matter produced in the Little Bang. }
\label{lilbang}
    \end{center}
\end{figure}

We are also seeing a strong correspondence with cosmology develop.  This is illustrated in
Fig. \ref{lilbang}.  The various stages of the evolution of matter in nuclear collisions is shown.
Before the collision, there is the Color Glass Condensate of the nuclear wavefunction.  At the moment of the collisions there is the initial singularity, analogous to that of the quantum gravity  and inflationary stages of cosmology, where quantum fluctuations are important.  After the singularity there
is a Glasma phase which has topological excitations analogous to those associated with
baryon number violation in electroweak theory. The Quark Gluon Plasma is analogous to both the
electroweak and QCD thermal phases of expansion in cosmology, and the deconfinement transition
is analogous to that which generates masses for electroweak bosons.

The bulk of this meeting concerns the properties and probes of the Quark Gluon Plasma.
Over the past few years we have seen a consensus develop that one is making a strongly 
interacting Quark Gluon Plasma at RHIC.  The range of work on this problem is enormous,
and I can only cover some highlights and more controversial aspects of the work on this problem.

I apologize to those whose work I cannot cover in this talk.  This is not due to lack of interest on 
my part, but the finiteness of space and time.  Also for reasons of space, all the citations in this talk
are either to presentations at this meeting or to reviews.  The citations to original literature may be found
in these papers.

\section{The  Initial Wavefunction}

\begin{figure}[ht]
    \begin{center}
        \includegraphics[width=0.70\textwidth]{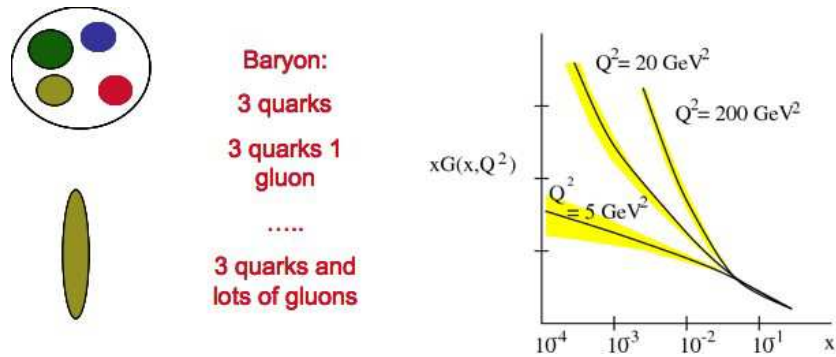}
        \caption{The leftmost figure on this slide illustrates the wavefunction for a gluon as a function
        of energy.  The rightmost figure illustrates the experimentally measured gluon distribution
        functions as a function of x. }
\label{wfn_glue}
\end{center}
\end{figure}

In the leftmost part of Fig. \ref{wfn_glue},  I show the  various Fock space components for
a baryon wavefunction.  At low energies, the dominant states for physical processes have three quarks and a few gluons.  In high energy collisions, many particles are produced.  These ultimately arise from 
components of the hadron wavefunction which have many gluons in them.  These components make
a gluon wall of longitudinal extents of $1/\Lambda_{QCD}$.  The density of gluons in the
transverse plane grows as the energy increases, as is also shown Fig. \ref{wfn_glue}

As described in the talks by Gelis and by Venugopalan, these gluons must become very tightly packed
together.  They form a high density, highly coherent condensate of gluons, the Color Glass Condensate (CGC).
Because the typical separation of gluons is small, at high enough energy $\alpha_S <<1$,
and the system is weakly coupled. Due to the coherence, it is also strongly interacting.
(An example of coherence greatly amplifying a very weak interactions is given by gravity.)
There is a typical momentum scale $Q_{sat}$, and gluons with momentum less
than this scale have maximal phase space density,
\be
	{{dN} \over {d^2p_Td^2r_Tdy}} \sim {1 \over \alpha_S}
\ee 
As one goes to higher energies, the saturation momentum increases, as more gluons are added to
the CGC.

The theory of the CGC has provided both a rich phenomenology as well as first principles understanding within QCD for the high energy limit. \cite{cgc} Some of the successes are described in the talk by Gelis and include
\begin{itemize}
\item{Geometric scaling in Deep Inelastic Scattering (DIS).}
\item{Diffractive DIS.}
\item{Shadowing of parton distributions.}
\item{Multiplicity distributions in hadron and nuclear collisions.}
\item{Limiting fragmentation.}
\item{Long range correlations for particle production.}
\item {Intuitive understanding of high energy limit for cross sections.}
\item{The origin of the Pomeron, Odderon and Reggeon}
\end{itemize}
This is a very active area of research with many new areas.  It has as one of its intellectual
goals, the unification of the description of all strong interaction processes at high energy
and as such involves different phenomena:  electron-hadron, hadron-hadron, hadron-nucleus and nucleus-nucleus collisions. 

\section{The Initial Singularity and the Glasma}

Before the collision of two hadrons, two sheets of Colored Glass approach one another.  Because
the phase space density of gluons is large, the gluons can be treated as classical fields.  The fields
are Lorentz boosted Coulomb fields, that is Lienard-Wiechart potentials, which are static in the
transverse plane of the hadrons and have $E \perp B \perp \hat{z}$, where $z$ is the direction of motion.\begin{figure}[ht]
    \begin{center}
        \includegraphics[width=0.50\textwidth]{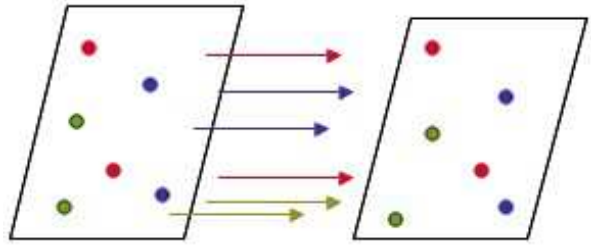}
        \caption{After the collision of two sheets of colored glass, longitudinal electric and
        magnetic fields associated with the Glasma form. }
\label{glasma}
\end{center}
\end{figure}
When the classical equations for the evolution of these fields is solved, in a very short 
time scale of order
$\Delta t \sim {1 \over Q_s} e^{-\kappa /\alpha_S}$, the fields change from purely transverse to purely
longitudinal.  This is because the collisions generate color electric and magnetic monopole charge
densities of opposite sign on the two sheets.  This description is reminiscent of the  flux tube models of
color electric fields which are used in phenomenological descriptions of low energy strong interactions.
At high energies,  there are both color electric and magnetic field because the electric and magnetic
fields were of equal magnitude in the CGC, and because of the electric-magnetic duality of QCD.

Fields with a non-zero $\vec{E} \cdot \vec{B}$ carry a topological charge.  In QCD, they are associated
with anomalous mass generation and chiral symmetry violation.  In electroweak theory, such
fields may be responsible for generating the baryon asymmetry of the universe.  In QCD, they may
generate the masses of those particles which constitute the visible matter of the universe. Each field
configuration violates CP.  An experimental discovery of the effects of such fields would be of great importance.
Theoretical ideas for experimental signatures are sketchy.\cite{cgc} 

The initial Glasma fields are unstable, and after a time scale of order $1/Q_s$,
instabilities begin to become of the order of the original classical fields, as described in the talks by Strickland.  (The origin of these Weibel instabilities was originally found
for plasmas close to thermal equilibrium by Mrowczinski, as described in Strikland's talk.)   Quantum fluctuations in the original
wavefunction can grow by these instabilities, and eventually overwhelm the longitudinal electric
and magnetic Glasma fields.  Perhaps these fields form a chaotic or turbulent liquid which might
thermalize and isotropize the system, as discussed by Asakawa. 

\begin{figure}[ht]
    \begin{center}
        \includegraphics[width=0.70\textwidth]{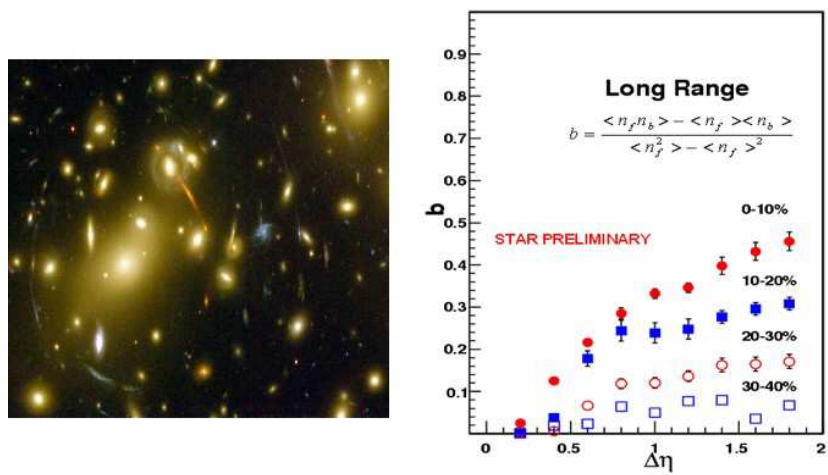}
        \caption{On the left is a photo of many galaxies.  On the right shows the long
        range forward backward correlations in rapidity measured in the Star experiment.
        Short range correlations have been subtracted from the plot shown. }
\label{seed}
\end{center}
\end{figure}

The amplification of quantum fluctuations to macroscopic magnitude is reminiscent of inflation
in the early universe.   These quantum fluctuations expand to size scale larger than the event
horizon during inflation, and are imprinted into the fabric of space-time.  When at much later times, the
event horizon size scale becomes of the order of galactic size scales, these fluctuations reappear,
and ultimately drive gravitationally unstable modes which form galaxies and clusters of galaxies.
In the leftside of Fig. \ref{seed}, a picture of distributions of galaxies is shown. 

In heavy ion collisions, analogous fluctuations in the hadronic wavefunction might
appear as rapidity dependent fluctuations.  On rapidity length scales larger than a few units, such
fluctuations would be frozen into the final state distribution of particles.  In a Star preliminary result presented in a poster session, the forward backward rapidity correlation was measured
as a function of centrality, as is shown in Fig. \ref{seed}.  This correlation is very strong and increases as a function of centrality
of the collision.  It can be understood arising from long range longitudinal fields, such as in the Glasma or string models, which reach maximum strength when the coupling becomes weak.

\section{The Strongly Interacting Quark Gluon Plasma}

Some of the first data from RHIC showed that there was strong radial and elliptic flow, and strong jet
quenching. \cite{whitepaper} The flow data could be described using perfect fluid hydrodynamic equations.  Although
there is a good deal of uncertainty due to lack of precise knowledge of the initial state,
final decoupling, equations of state and the magnitude of viscous corrections, it is nevertheless
remarkable that  perfect fluid hydrodynamic computations can describe the data. 
Such a description was not possible for data at lower energies. \cite{whitetheory}

\begin{figure}[ht]
    \begin{center}
        \includegraphics[width=0.80\textwidth]{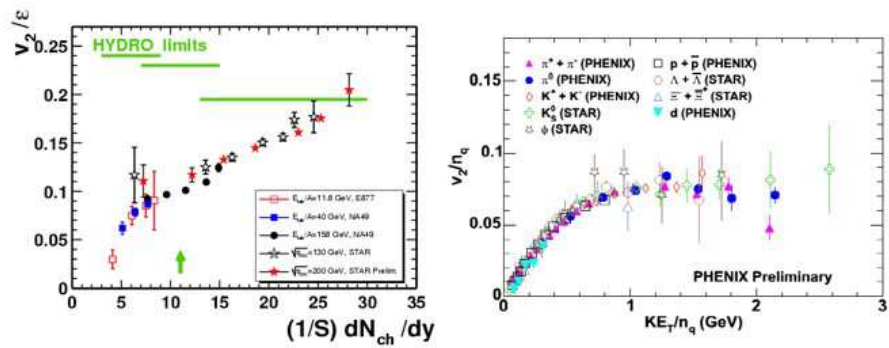}
        \caption{The left hand figure shows elliptic flow scaled by the eccentricity of the overlapping region of colliding nuclei as a function of the number of produced particles per unit area.  The right hand
        scales the flow of individual particles by the number of valence quark constituents verses
        the transverse kinetic energy scaled by the same factor.  }
\label{flow}
\end{center}
\end{figure}

There were many plots
in this Quark Matter (and previous ones) which showed the agreement between hydrodynamic computations
and distributions of produced particles in $p_T$ and $y$, summarized in the talks by Barranikova and Nonaka .  The agreement between data and experiment
seems to be improved by the more careful treatments of late time evolution in the form of hadrons,
and using Color Glass initial conditions.\cite{whitetheory}  In Fig. \ref{flow},
I show the comparison of elliptic flow scaled by eccentricity of the colliding region vs
the  mutliplicity of produce particles per unit area..  The data seems to saturate hydrodynamic bounds using zero
viscosity and Glauber type initial conditions.  The upper bound would be somewhat higher if
Color Glass initial conditions were used.  In the second plot of Fig. \ref{flow}, the elliptic flow is scaled by the number of quark constituents of produced particles as a function of transverse energy scaled
by the same factor.  This is predicted in coalescence models.  The agreement is extraordinary, given that
coalescence models do not have simultaneous energy and momentum conservation.  
(Worse things can happen than having a theoretical computation agree too well with experiment.)
Such coalescence models provides reasonable support for  the idea that the matter in early times was composed of quarks and gluons.  

Jet quenching computations have  been described in the talk by Majumder and many parallel session
talks.  The suppression by a factor of four out to transverse momenta of about $20~GeV$, corresponding to several $GeV$ energy loss is stunning and was not anticipated.  There are a variety of different
QCD based models which adequately describe the suppression of light mass hadrons.

These observations have led to a consensus that one has produced a strongly interacting Quark
Gluon Plasma, and that to a fair or good or perhaps excellent approximation, the system is thermalized.
A well thermalized system is a perfect fluid.  The outstanding issue is how perfect is the perfect fluid.

This may be parameterized by the dimensionless ratio of viscosity to entropy density, $\eta/s$.
As argued in Son's and Liu's talks,   $N=4$ supersymmetric
Yang-Mills theory  satisfies a bound
\be
	{\eta \over s} \ge {1 \over {4\pi}}
\ee	
with the lower bound approached for infinite coupling.
This is a beautiful result which leads to new insight.  Although $N=4$ supersymmetric theories bear
no direct relation to our physical world, this bound was also more generally argued as 
a consequence that in the strong coupling limit of QCD,  the limiting small
viscosity is reached when  the mean free path of a particle become equal to its deBroglie wavelength.  
Below
this limit, scattering is difficult due to quantum mechanical coherence. 

How close we are to the limit $\eta /s = 1/4\pi$? Kapusta argued that for a wide class of liquids,
$\eta/s$ has a minimum close to a critical point.  In the QGP, this ratio is an increasing
function of increasing temperature within perturbative computation.  At low temperatures
in a pion gas, it is an increasing function of decreasing temperature, so it is plausible that it has a minimum at the transition.  (For a first order transition there would be a minimum
contiguous to a discontinuity.)

Is $\eta/s$ close to the minimum value for temperatures accessible at RHIC in the QGP?  
There are reasons to be skeptical: (1)  In this meeting there were new computations which argued that one does not need extremely large cross sections within transport computations to describe the elliptic flow data, as shown by Ko.   (2) Flow results depend on initial conditions, about which there is not yet
consensus, the way that the equation of state is treated, and how one coalesces quarks and gluons into
hadrons.  There is much work now on initial conditions,
so this has potential for resolution.  If the initial conditions are more like those predicted from the
Color Glass Condensate, one generates more elliptic flow, which could be reduced by increasing the viscosity. (3)
We saw at this meeting the first realistic computations with
relativistic hydro and viscosity, and soon there should be computations of elliptic flow.  The preliminary
results from these hydro computations indicate that varying viscosity can be compensated
by various other uncertainties in the computations.

One test of the effect of viscosity is given by the left hand plot of Fig. \ref{flow}.
At the increased energy of LHC, the number of produced particles per unit area is larger.  If one "exceeds the hydro bounds", then one concludes that the Glauber based initial conditions at RHIC energy
were not correct and were compensated by viscosity. 

\section{Jets as Probes of the Quark Gluon Plasma}

Jets have been much discussed as probes of the matter produced in heavy ion collisions.
As discussed by Majumder, there are a variety of model computations which are QCD based,
and have varying degrees of fidelity with the underlying theory.  One of the problems
is the low momentum exchanges in scattering, where QCD perturbation theory is not applicable.
Another is that the jet energy loss involves a time integral over the collision and is order one
sensitive to the matter when it is at low density.  Even with these caveats, the work which has been
done to analyze RHIC collisions is very impressive.  

I will concentrate on two issues in this talk.  The first is the possible observance of a Mach cone.
The second is the mystery of charm energy loss.

\begin{figure}[ht]
    \begin{center}
        \includegraphics[width=0.80\textwidth]{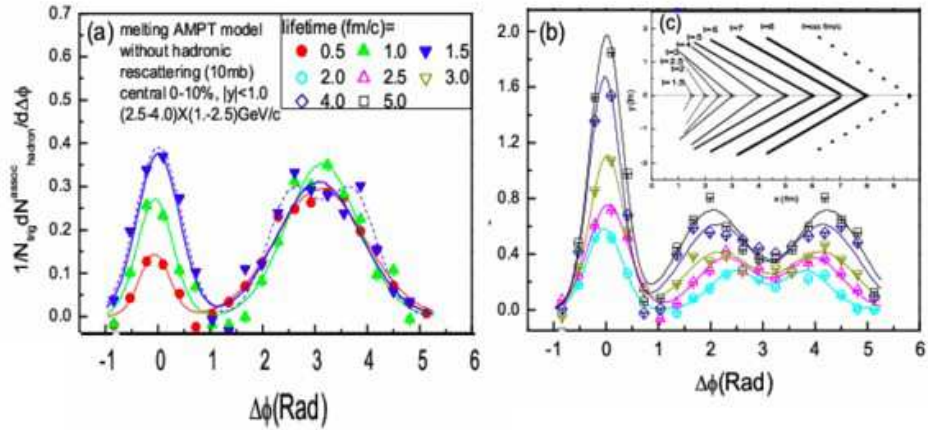}
        \caption{The distribution of particles associated with a high $p_T$ primary as a function
        of azimuthal angle for various upper cutoffs on the energy of the associated particle. }
\label{mach}
\end{center}
\end{figure}

The idea of a Mach cone in heavy ion collisions has origins in the Frankfurt group, who a long time ago suggested Mach cones may appear in ultrarelativistic hadron-nucleus and nucleus-nucleus collisions..  
In RHIC collisions, if one tags a jet by measuring a high $p_T$ particle, one can look at azimuthal angles near $\pi$ radians for the associated jet.  As discussed by Wang and by Solana,  one finds the high momentum fragments of the away side jet  are suppressed, a fact  which
is consistent with the jet quenching found by direct measurement.  If one looks at lower momentum for the associated jet, it reappears, but its distribution is widened, and seems to have a two peak 
structure, as shown in Fig. \ref{mach}  One might naively expect such a structure for either a Mach cone
or Cerenkov radiation.  There are problems with either of these interpretations since the putative cone has such a wide opening angle.  This would required very small sound velocities for a Mach cone interpretation, $v_s^2 \le 10^{-2}$, as pointed out by Stoecker.  It may be possible that this is simply due
to multiple scattering including the effects of Sudakov form factors, as discussed by Salgado.  
In multiple scattering scenarios, one does not expect a conical structure, and preliminary analysis
of three particle correlations favors  a cone.  It is clearly very early to make a firm conclusion as to the origin of this effect, as the devil may be in the details of both the theoretical and experimental
analysis.
 \begin{figure}[ht]
    \begin{center}
        \includegraphics[width=0.50\textwidth]{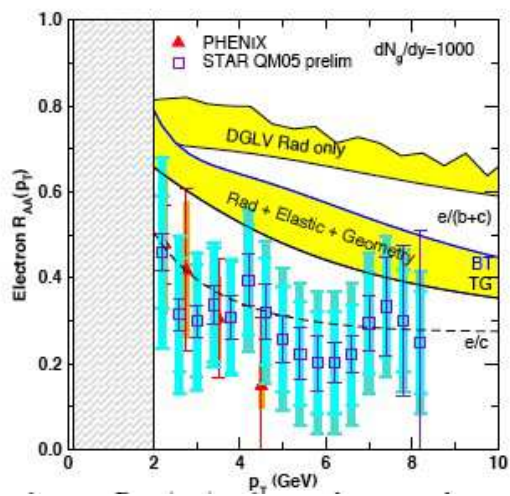}
        \caption{The ratio $R_{AA} (p_T)$ for unaccompanied electrons.}
\label{charm}
\end{center}
\end{figure}

In Fig \ref{charm}, I present the results of the Star and Phenix collaborations for the ratio
$R_{AA} (p_T)$ of unaccompanied electrons.  This can be related to the quenching factor for
charm  and bottom quarks.  These results make big problems for theorists.  Heavy quarks should have less energy loss than light quarks.  To see this, imagine a heavy quark is scattered in its
rest frame.  It gets excited and decays with a typical QCD energy scale and its fractional
recoil energy  is therefore suppressed by $1/M_{charm}$.  Boosting to a fast moving frame, the energy loss
of a fast moving quark is suppressed by the same factor, and is the origin of the dead cone effect of Dokshitzer and Kharzeev.\cite{whitetheory}  On the other hand, the data indicate that charm quarks
lose their energy like light quarks. There is moreover some reasonable fraction of bottom quarks 
in the data, which should lose even less energy.  Radiative energy loss computations cannot reproduce the data, and elastic scattering energy loss helps somewhat but not enough.
One should comment that these first principle computations are hard, and are not strictly perturbative,
but nevertheless, the level of disagreement between theory and experiment is not reconciled.

\section{Strings and AdSCFT}

There has been some discussion at this meeting about string theory and its relationship to heavy ion 
physics.  AdSCFT is a mathematical trick which allows one to compute the properties of $N=4$ supersymmetric Yang-Mills theory in terms of gravitational theory in curved space.  
The strong coupling limit of N=4 SUSY corresponds to weak coupling gravity, allowing strong
coupling computations for the Yang-Mills theory.
 
N= 4 supersymmetric Yang-Mills is not QCD.  In the very nice talk by Liu, he cautioned that:
\begin{itemize}
\item{ It has no mass scale and is conformally invariant.}
\item{It has no confinement and no running coupling constant.}
\item{It is supersymmetric.}
\item{It has no chiral symmetry breaking or mass generation.}
\item{It has six scalar and fermions in the adjoint representation.}
\end{itemize}
The interesting applications of this correspondence for our field is in the strong
coupling limit. 
Even in lowest order strong coupling computations it is very speculative to make relationships
between this theory and QCD, because of the above.  It is much more difficult to relate
non-leading computations to QCD.  

It may be possible to correct some or all of the above problems, or, for various physical problems,
some of the objections may not be relevant.  As yet there is not consensus nor compelling arguments
for the conjectured fixes or phenomena which would insure that the $N=4$ supersymmetric Yang Mills results
would reliably reflect  QCD.

Further, these computations are applied to the QGP at temperatures above the deconfinement 
transition temperature, where the validity of
strong coupling limit is arguable.  Lattice computations give reliable computations of the 
properties of the QGP, and indicate the coupling is of intermediate to weak strength in this region,
as discussed by Hatsuda. There are improved weak coupling computations which agree
with lattice data at temperatures more than a few times $T_c$, as discussed by Blaizot.

The AdSCFT correspondence, is probably best thought of as a discovery tool with limited
resolving power.  An example is the $\eta/s$ computation.  The discovery of the bound on $\eta/s$
could be argued be verified by an independent argument, as
a consequence of  the deBroglie wavelength of particles becoming of the order of mean free paths.  
It is a theoretical discovery but its direct applicability to heavy ion collisions remains to be shown.

Certainly,  computations in lattice gauge theory
or in weak coupling provide reliable QCD computations of a limited set of
physical phenomena.  One has  had  hope
for analytic methods in  the strong coupling dynamics of QCD, and we should continue to have such hope.  The advocates of AdSCFT correspondence
are shameless enthusiasts, and this is not a bad thing. Any 
theoretical physicist who is not, is surely in the wrong field.  Such enthusiasm will hopefully be balanced by commensurate skepticism.

\section{Acknowledgments}
I gratefully acknowledge conversations with 
Rob Pisarski, Dima Kharzeev, Nick Samios,  Raju
Venugopalan  and Urs Wiedemann who provided input on the subject of this talk.  

This manuscript has been authorized under Contract No. DE-AC02-98CH0886 
with
the U. S. Department of Energy.


\section*{References}
\begin{thebibliography} {10}

\bibitem{cgc} For a review of the Color Glass Condensate, see E. Iancu and R. Venugopalan,
in QGP3 edited by R. Hwa and X. N. Wang, p 249.

\bibitem{whitepaper}  Reports of the Star, Phenix, Phobos and Brahms experimental collaborations
 in {\it Nucl. Phys} {\bf A757} (2005)
 Brahms Collaboration p  1;  Phobos Collaboration p 28; Star Collaboration p 102; Phenix Collaboration 184.


\bibitem{whitetheory} For a theoretical review of the properties of the QGP and it effects in heavy ion
collisions see Workshop on New Discoveries at RHIC: The Current Case for
the Strongly Interactive  Quark Gluon Plasma, Brookhaven, May 14-15 , 2004,  {\it Nucl. Phys.} {\bf A750} (2005) .



\end{thebibliography}
\end{document}